\documentclass{ws-ijmpd}
\usepackage[super,compress]{cite}

\usepackage{amsmath}
\usepackage{graphics}
\usepackage{hyperref}
\usepackage{multirow}
\usepackage{caption}

\usepackage{color,soul}
\usepackage{csquotes}

\def\be{\begin{equation}}
\def\ee{\end{equation}}
\def\bea{\begin{eqnarray}}
\def\eea{\end{eqnarray}}
\def\bn{\begin{enumerate}}
	\def\en{\end{enumerate}}
\def\bsube{\begin{subequations}}
	\def\esube{\end{subequations}}

\def\nn{\nonumber}
\def\ll{\left}
\def\rr{\right}
\def\mc{\mathcal}
\def\mb{\mathbf}
\def\Lc{\mc{L}}
\def\FAP{{FAP\,}}
\def\DP{{ DP\,}}
\def\Ro{\boldsymbol{\rho}}
\def\Mpc{{\text{Mpc}~}}
\def\Gpc{{\text{Gpc}~}}

\def\z{\mb{z}}

\def\h{\mb{h}}
\def\f{\text{F}}
\def\F{\mb{F}}
\def\Ro{\boldsymbol{\rho}}

\def\Mpc{\text{Mpc}}

\begin{document}

\markboth{K Haris, Vinaya Valsan and Archana Pai}
{Performance of multi-detector hybrid statistic in targeted CBC search}

%
%

\title{Performance of multi-detector hybrid statistic in targeted compact binary coalescence search}

\author{K Haris}

\address{International Centre for Theoretical Sciences, \\ Tata Institute of Fundamental Research, \\ Bangalore 560089, India\\
	haris.mk@icts.res.in}

\author{Vinaya Valsan}

\address{Indian Institute of Science Education and Research Thiruvananthapuram, \\
	College of Engineering Campus,   Thiruvananthapuram, \\ Kerala	695016, India\\
	vinayavalsan2010@iisertvm.ac.in}

\author{Archana Pai}

\address{Department of Physics,  Indian Institute of Technology Bombay, \\ Powai, Mumbai, Maharashtra 400076, India \\ \vspace{0.2cm} Indian Institute of Science Education and Research Thiruvananthapuram, \\
	College of Engineering Campus,  Thiruvananthapuram, \\ Kerala	695016, India\\
	archana@phy.iitb.ac.in}

\maketitle
\begin{history}
\received{---}
\revised{---}
\end{history}
\begin{abstract}
	In this paper we compare the  performance of two  likelihood ratio based detection  statistics namely maximum likelihood ratio  statistic and {\it hybrid} statistic designed for the  detection of gravitational waves from compact binary coalescence using multiple interferometric  detector networks. We perform simulations with  non-spinning double neutron star binary system and neutron star - black hole binary systems with spinning as well as non-spinning black hole component masses. The binary injections are  distributed uniformly  in volume up to 1 Gpc.  We observe that, on  average, the  maximum likelihood ratio  statistic  recovers $\sim 34.45\%$,  $\sim 49.69\%$, $\sim 61.25\%$ and $\sim 69.67\%$  of injections in 2, 3, 4 and 5 detector networks respectively in the case of  neutron star - black hole injections for a fixed false alarm probability of  $10^{-7}$ in Gaussian noise. Further, we note that, compared to the  maximum likelihood ratio   statistic, the   {\it hybrid} statistic recovers  $\sim 7.45\%$, $\sim 4.57\%$, $\sim 2.56\%$ and $\sim 1.22\%$ more injections in 2, 3, 4 and 5 detector networks respectively for the same false alarm probability in Gaussian noise. On the other hand, among binary neutron star injections,  the  maximum likelihood ratio  statistic  recovers  $\sim 5.587\%$,  $\sim 9.917\%$, $\sim 14.73\%$ and $\sim 19.86\%$ of injections   in 2, 3, 4 and 5 detector networks respectively and the   {\it hybrid} statistic recovers $\sim 14.63\%$, $\sim 12.91\%$, $\sim 11.49\%$ and $\sim 10.29\%$ more injections compared to maximum likelihood ratio  statistic  in 2, 3, 4 and 5 detector networks respectively.
\end{abstract}

\keywords{Gravitational waves; Multi-detector search; Compact binary coalescence.}

\ccode{PACS numbers:04.80.Nn, 07.05.Kf, 95.55.Ym}


\section{Introduction}
\label{Introduction}
Detection of gravitational waves (GW) from  two binary black hole mergers  \cite{PhysRevLett.116.061102, PhysRevLett.116.241103,PhysRevLett.118.221101} by Advanced LIGO detectors has opened a new observational window to our universe.
The Virgo interferometer \cite{TheVirgo:2014hva,Avirgo}  joined the  LIGO detectors  and made first double neutron star (DNS) observation in August 2017 \cite{TheLIGOScientific:2017qsa}.
Addition of more advanced interferometers like KAGRA \cite{PhysRevD.88.043007,0264-9381-29-12-124007} and LIGO-India \cite{LIGOIndia} located in different continents will provide additional information like source location and polarization of the incoming GW signal \cite{Tagoshi:2014xsa}. GW from inspiraling compact binary coalescences (CBC) composed of neutron stars (NS) and black holes (BH) are primary sources for interferometric multi-detector  GW networks. Based on LIGO's initial observations,   it is expected that the Advanced LIGO detectors could  observe tens of  DNS and NS-BH merger events per year  along with  hundreds of binary BH mergers once they achieve their designed sensitivity \cite{TheLIGOScientific:2016pea,Abbott:2016nhf,Abbott:2016ymx,Abadie:2010cf}. 

The coherent detection  schemes for the GW search of CBC sources with known source location (targeted search) has been developed in literature by various groups \cite{Pai:2000zt,Harry:2010fr,Haris:2014fxa}. The two stream  maximum likelihood ratio (MLR) detection statistic is obtained by combining data  from different detectors in a phase coherent fashion. The coherent  scheme has been implemented as a targeted GW follow-up search of gamma ray bursts (GRBs) observed by the Inter Planetary Network (IPN) satellites, where the source location is known a priori from IPN data \cite{Abadie2012,PhysRevLett.113.011102}. Further, in \cite{Macleod2016} the authors have designed all-sky coherent search scheme for LIGO network as an extension of MLR based targeted coherent analysis by placing templates in the time delay coordinates \cite{0905.4832}. 

In \cite{Haris:2015iin}, authors developed a new {\it hybrid} statistic for the coherent search of CBC sources.
The Hybrid statistic is a single stream detection statistic and is defined as the maximum of MLR statistic tuned for face-on/off binaries. 
The noise in the synthetic data streams increases the false alarm probability (FAP) in the two stream MLR statistic. Owing to the single stream, the hybrid statistic has low FAP. The low FAP sets low threshold and allows to probe deeper in noise with little compromise on the signal to noise ratio (SNR) and hence on the detection probability (DP).
Though suboptimal, \textit{hybrid} statistic can recover more than $98\%$ of the optimum SNR for a wide range of inclination angles ($\epsilon<70^{\circ}$ and $\epsilon>110^{\circ}$) for non-spinning binary systems \cite{Haris:2015iin}.
The work clearly demonstrates (both analytically  and numerically ) that the {\it hybrid} statistic  can recover more CBC events compared to the MLR statistic for a given type of source (due to the increased distance reach) for a typical case of 3 detector network.  

In this work, we carry out extensive performance comparison of MLR statistic and the {\it hybrid} statistic for all the possible two, three, four and five detector networks  with non-spinning DNS binary injections and non-spinning as well as aligned spin NS-BH binary injections. The sources are arbitrarily located as well as oriented in the sky in a distance range of ($100 \ \Mpc - 1 \ \Gpc$) distributed uniformly in volume. We demonstrate that the {\it hybrid} statistic  gives on an average $\sim 14.63\%$ ($7.45\%$), $\sim 12.91\%$ ($4.57\%$), $\sim 11.49\%$ ($2.56\%$) and $\sim 10.29\%$ ($1.22\%$) relative improvement in the injection recovery  rate over  MLR statistic  respectively with 2, 3, 4 and 5 detector combinations for a fixed FAP of $10^{-7}$ with DNS (NS-BH) binary injections. By design the {\it hybrid} statistic gives best performance for near face-on/off injections and gives worst performance for near edge-on injections.

The paper is organized as follows. Section \ref{globalnetwork} gives an overview of  the global network of interferometers.  The properties of various multi-detector networks are discussed in Section \ref{SC}. In Section \ref{statistics}, we briefly  discuss the two detection statistics used for multi-detector search of CBC signals namely MLR statistic and {\it hybrid} statistic.  In Section \ref{simulations}, we carry out numerical simulations to estimate detection efficiency of MLR statistic as well as {\it hybrid} statistic for various detector networks. Finally, in Section \ref{conclusions}, we summarize the results.
\section{Ground based GW multi-detector network}\label{globalnetwork}
A global network of ground-based advanced optical laser interferometric detectors will be complete in a decade. The 4km arm-length twin Advanced LIGO detectors  are located at  Hanford (H) and Livingston (L)  in the USA \cite{0264-9381-32-7-074001,Harry:2010zz,PhysRevLett.116.131103, Abbott:2016jsd}. The first observation run of Advanced LIGO detectors took place for a period of  four months starting from September 2015. During the run, the LIGO detectors made the historical detection of GW from three binary BH mergers namely GW150914 (Sept.14, 2015), GW151226 (Dec.26, 2015), GW170104 (Jan.04, 2017) \cite{PhysRevLett.116.061102, PhysRevLett.116.241103,PhysRevLett.118.221101,TheLIGOScientific:2016wfe}.The second observation run is ongoing. The $3$km Virgo (French-Italian) detector (V) located in Pisa, Italy has joined the second observation run \cite{TheVirgo:2014hva,Avirgo}. The Japanese  cryogenic detector KAGRA (K) located in 	Kamioka Observatory, Gifu  is a $3$km detector \cite{PhysRevD.88.043007,0264-9381-29-12-124007} and is expected to join the global network within a decade. The Indo-US $4$km LIGO detector in India (I)\footnote{In this study, we take Pune, India as the location for the detector in India.}, an extension of LIGO network is expected to join the network in a decade \cite{LIGOIndia}. These five advanced interferometric detectors will together form a ground based $km$ arm-length Michelson type (with perpendicular arms) 5-detector network, which we use in this work. The orientations and locations of the detectors are taken from Table 1 of \cite{Schutz:2011tw}.

\subsection{Inspiral Signal in a Multi-detector network}

The strain measured by the m-th gravitational detector in a network of detectors is given by,
\begin{equation}\label{signal}
s_m(t) \equiv \f_{+m} h_{+}(t)+\f_{\times m}h_{\times}(t),
\end{equation}
where $h_+$ and $h_{\times}$ are the two polarizations of GW in Einstein's gravity and are functions of binary component masses, distance to the source from the observer and inclination angle ($\epsilon$). 
GW polarizations in frequency domain are given by,
\begin{eqnarray}
\tilde{h}_+(f)&=& A(m_1,m_2,\chi,r) \frac{1+\cos^2\epsilon}{2} \tilde{h}_0(f;m_1,m_2,\chi) e^{i \phi_a}\\
\tilde{h}_{\times}(f)&=& A(m_1,m_2,\chi,r)  \cos \epsilon \, \tilde{h}_{\pi/2}(f;m_1,m_2,\chi) e^{i \phi_a}
\end{eqnarray}	
where $\tilde h_0(f) =  i \tilde h_{\pi/2}(f) \equiv f^{-7/6} e^{i \varphi(f;m_1,m_2,\chi)}$ (for $f>0$) is the frequency dependence of the signal, with  the restricted 3.5 PN phase  $\varphi(f)$  \cite{PhysRevD.84.084037}. The spin of BH is quantified in terms of a dimensionless number $\chi=||\frac{c \vec{S} }{G m_1^2}||$ where $\vec{S}=\{S_x,S_y,S_z\}$ is the angular momentum vector of the BH component.  Here, $\f_{+m}(\theta,\phi,\psi,\alpha_m,\beta_m,\gamma_m)$ and $\f_{\times m}(\theta,\phi,\psi,\alpha_m,\beta_m,\gamma_m)$ are the  antenna pattern functions of $m$-th detector, which give the directional response of an individual interferometer defined with respect to a reference frame attached to the center of Earth, called the Earth's frame  \cite{RefFrames} (For detailed expressions, See Equation.7 of \cite{Haris:2014fxa}). The ($\theta ,\phi$) is the binary source location with respect to this frame. The detector frame is the frame attached to the individual detector such that the z-axis of the frame is directed towards the local zenith of the detector and the x-y plane contain the detector arms.  The orientation of detector frame with respect to the Earth's frame gives the orientation of the detector ($\alpha_m,\beta_m,\gamma_m$). The wave frame is the frame attached with the incoming GW and its x-y plane is perpendicular to the line of sight to the binary from Earth's center. The orientation of the binary plane in this frame gives the polarization angle $\psi$. The inclination of the binary angular momentum vector with the line of sight  gives  $\epsilon$. 

\subsection{Signal to noise ratio and Skymap of multi-detector networks}
For a global multi-detector network with independent noises, the network SNR square $\Ro^2$ is the sum of the individual SNR squares and is 
given by\footnote{The scalar product
	$(\mathbf{a} |\mathbf{b})_m =  4 \Re [ \int_0^{\infty} { \tilde{\tilde a}}(f) ~ {\tilde b}^*(f)~  df],$ 
	where  $ \tilde{\tilde a}(f) = \tilde a(f) / S_n(f)$ is the over-whitened version of frequency series $\tilde a(f)$ with $S_n(f)$ as the one-sided noise power spectral density (PSD) of $m$-th detector.}
\begin{equation}
\Ro^2 = \sum_{m} \rho_m^2 \equiv \sum_m \f_{+m}^2 (\h_+|\h_+)_m + \f_{\times m}^2 (\h_\times | \h_\times)_m  \,.\label{SNR}
\end{equation}

\begin{figure*}[h]
	\centering
	\includegraphics[width=\textwidth]{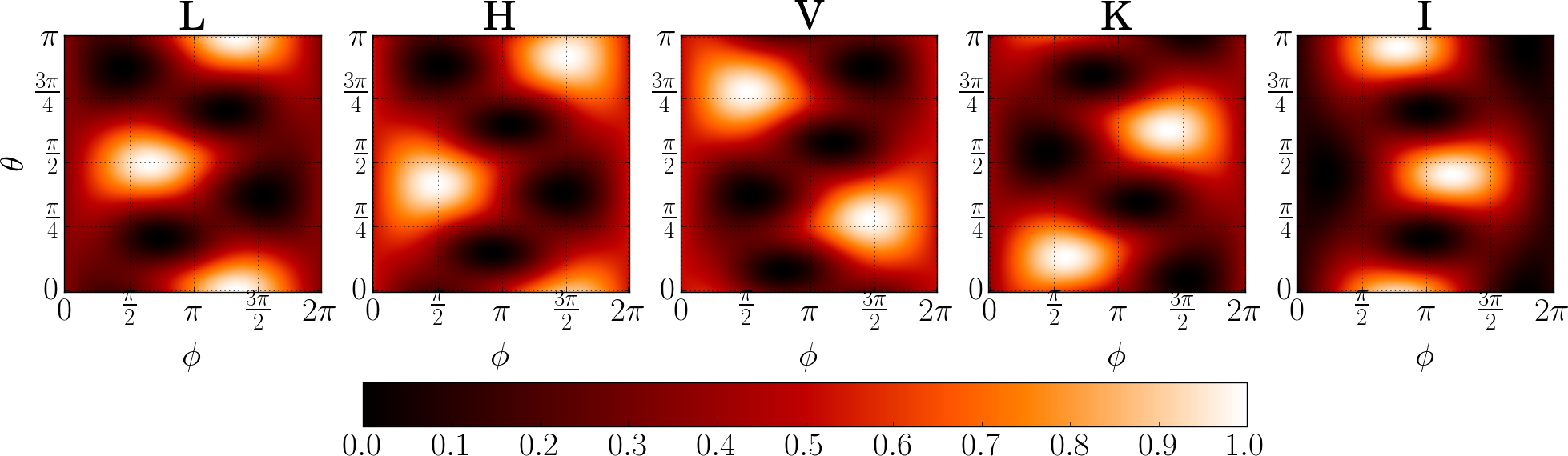} 
	\caption{Sky map of antenna pattern function ($\f_{+m}^2 + \f_{\times m}^2$) for individual detectors. Each interferometer has 2 maxima along the local zenith and beneath and four blind directions along the perpendicular bisectors of the interferometer arms denoted by the dark patches.}\label{Chap3.antennapattern}
\end{figure*}
\begin{figure*}
	\centering
	\includegraphics[width=\textwidth]{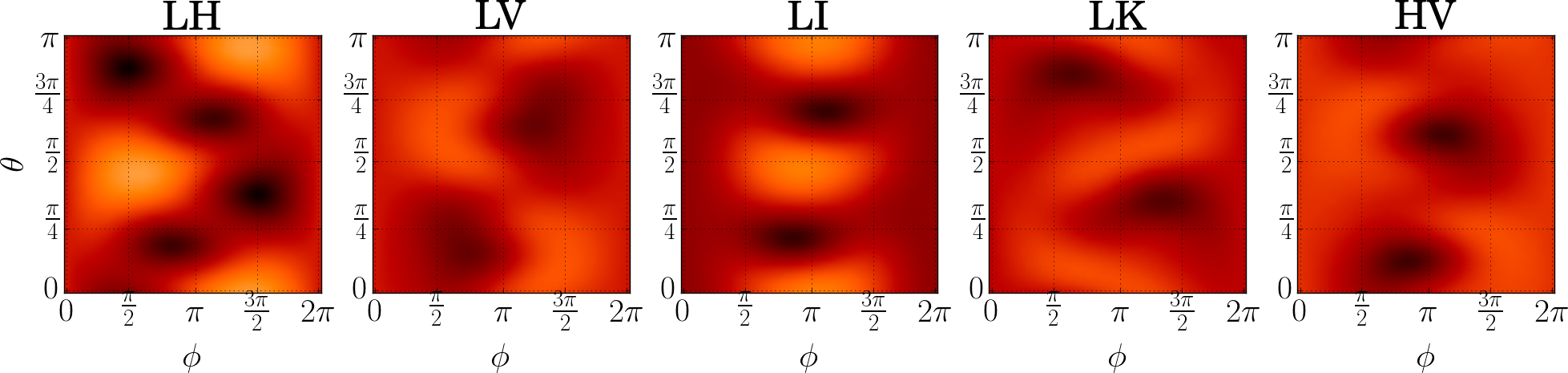}\vspace{-0.34cm}\\
	\includegraphics[width=\textwidth]{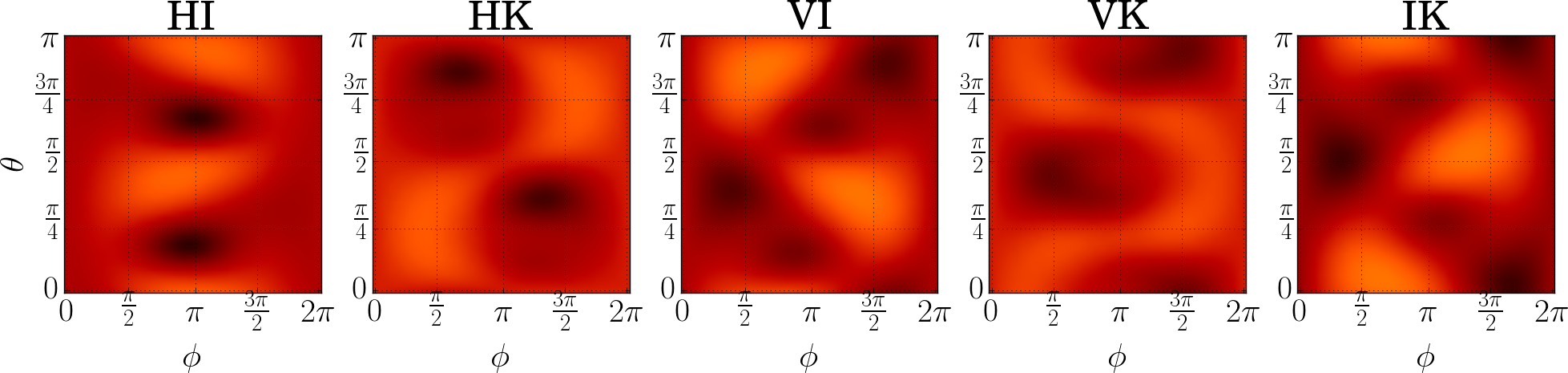}\vspace{-0.34cm}\\	
	\includegraphics[width=\textwidth]{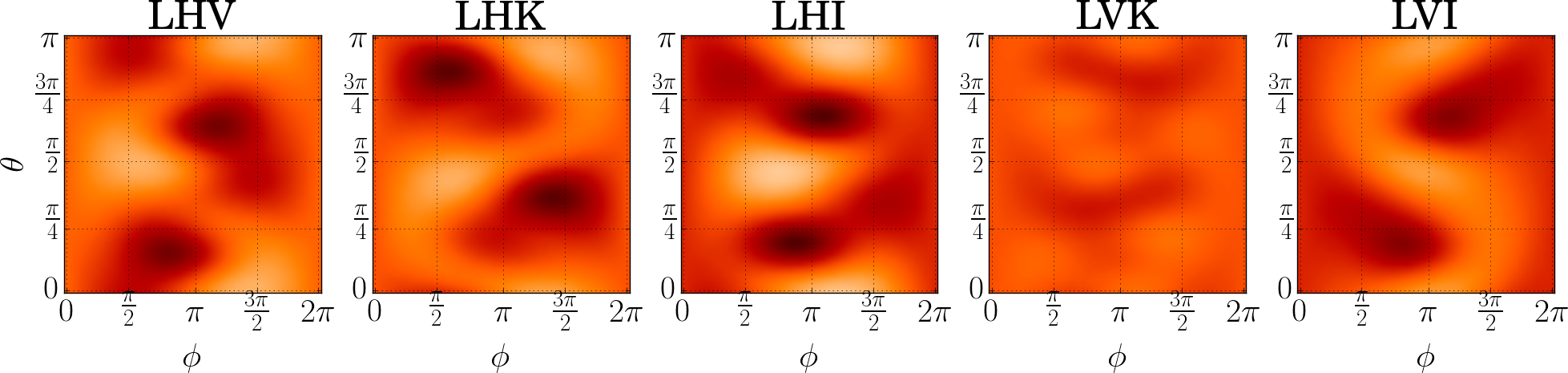}\vspace{-0.34cm}\\
	\includegraphics[width=\textwidth]{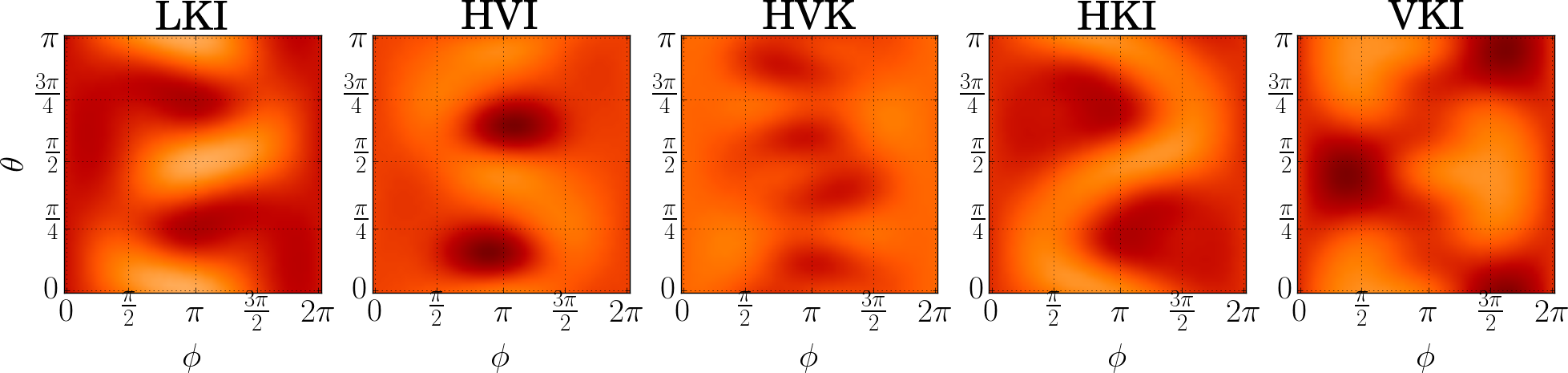}\vspace{-0.34cm}\\
	\includegraphics[width=\textwidth]{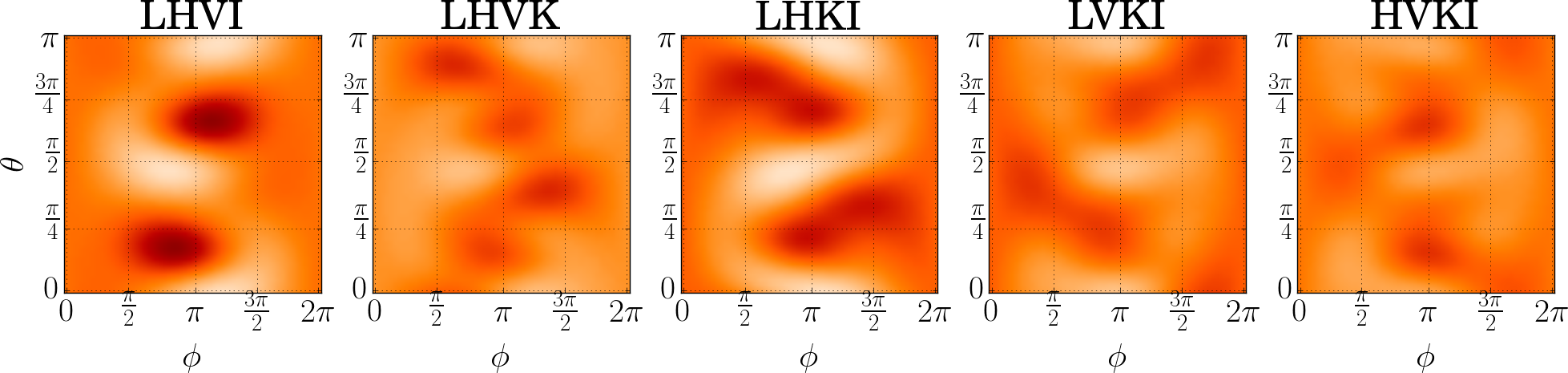}\vspace{-0.0cm}\\
	\includegraphics[width=\textwidth]{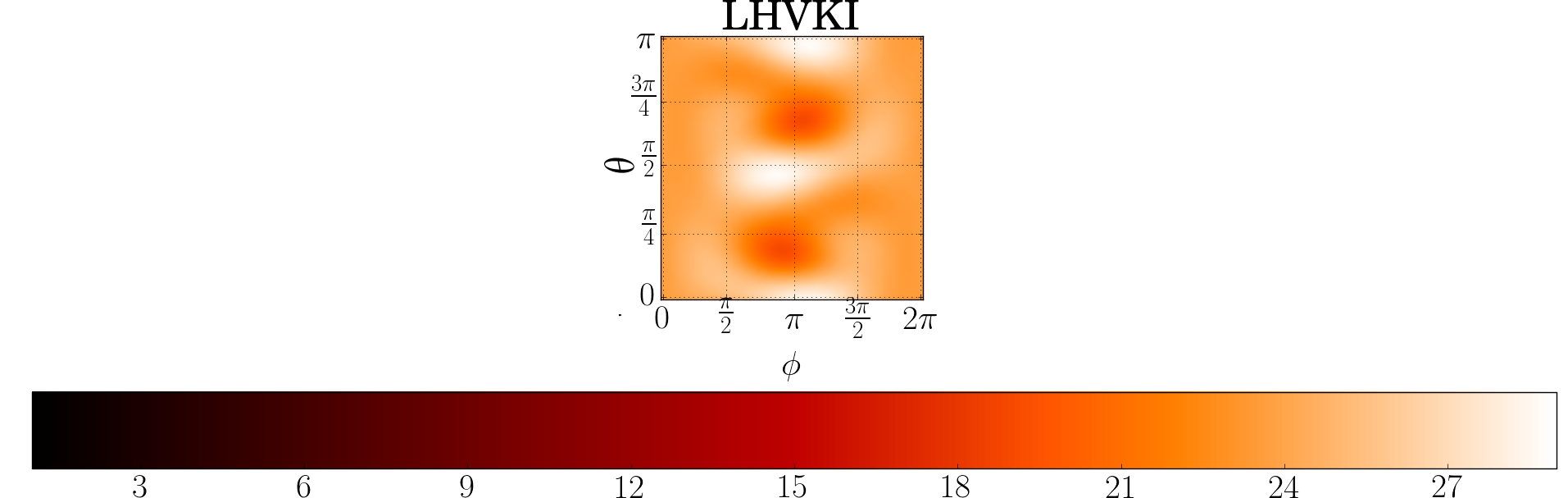}
	\caption{ \label{Chap3.SkyCoverage1} Sky map of network  SNR $\Ro$ of inspiral signal with masses ~$(1.4-10)M_\odot$, inclination angle $\epsilon=0$, polarization angle $\Psi =0,~$ and distance $r= 500 \ \Mpc$ for various 2, 3, 4 and 5 detector network configurations. We assume all the detectors with \enquote{zero-detuning, high power} Advanced LIGO noise curve \cite{aLIGOSensitivity}.}
\end{figure*}
In Figure \ref{Chap3.antennapattern}, we draw the sky maps, i.e ($\f_{+m}^2 + \f_{\times m}^2$) for the individual detectors. The antenna pattern for each of the GW interferometers are quadrupolar in nature with 2 maxima (along the local zenith and beneath) shown in white color and 4 blind directions in the plane of the
interferometer (directions  along the bisectors of arms) shown in black color. Due to the distinct geographical locations, the planes of each
of these detectors are distinct and hence the maximum response and the blind directions as well (See Fig \ref{Chap3.antennapattern}). Please note that the two maxima of the detector V overlap with the two distinct blind spots of L and H respectively
filling the blind directions. Thus, we expect that, when more detectors start taking the data together, more sky becomes visible to the multi-detector network as a whole. Further, with improved visibility, the network sees much deeper in space.

In Figure \ref{Chap3.SkyCoverage1} we draw the directional SNR for various 2, 3, 4 and  5 detector networks with LIGO Hanford (H), LIGO Livingston (L), Virgo (V), KAGRA (K) and LIGO India (I). For this exercise, we assume all detectors to have \enquote{zero-detuning, high power} Advanced LIGO noise PSD \cite{aLIGOSensitivity}. We consider the source to be face-on non-spinning NS-BH with masses $(1.4-10)M_{\odot}$ located at 500 Mpc.
Broadly we can see that, with more number of  detectors, more and more sky becomes brighter, {\it i.e.,} sensitive to a network. In Figure \ref{Chap3.SkyCoverage1}, most sensitive regions are white and least sensitive regions are black in color.  

\begin{figure*}[h]
	\centering
	\includegraphics[scale=0.35]{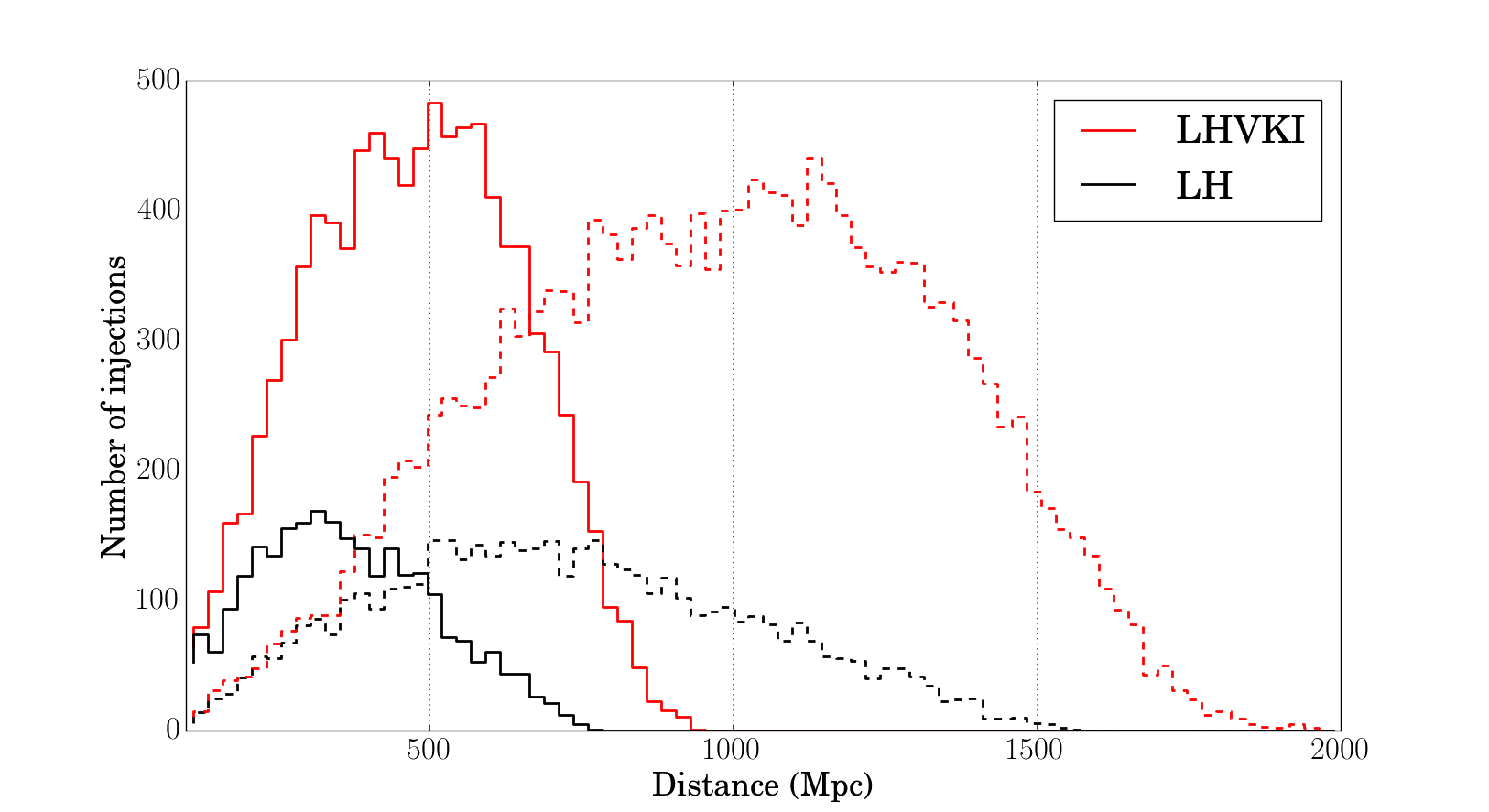} 
	\caption{Distance distribution of binary injections with SNR $ >6$ for two detector combinations: LH and LHVKI with DNS systems(solid line) of mass $(1.4-1.4)M_{\odot}$ and NS-BH systems (dashed line) of mass $(1.4-10)M_{\odot}$.}\label{HIST_dist_reach}
\end{figure*}

In Figure \ref{HIST_dist_reach} we plot the distribution of binary injections with SNR $>$6 for two different detector combinations: LH and LHVKI for both DNS and NS-BH systems. We distribute DNS injections of mass $(1.4-1.4) M_{\odot}$ and NS-BH injections of mass $(1.4-10)M_{\odot}$ uniformly in volume with the binary orientation $\cos \epsilon$ and $\psi$ sampled uniformly from the range $(-1,1)$ and $(0^{\circ},360^{\circ})$ respectively. It can be seen that the number of recovered injections are maximum around $500 \ \Mpc$ for DNS system and $1 \ \Gpc$ for NS-BH systems in the case of detector network LHVKI. This distance is shifted to  around $300 \ \Mpc$ for DNS and $700 \ \Mpc$ for NS-BH in the case of  network LH. Owing to low mass of DNS system, the luminosity distance of the farthest recovered DNS injection  is smaller than that of  farthest recovered NS-BH injection.
In doing the simulations in Section\ref{simulations}, we distribute DNS and NS-BH injections uniformly in volume within a distance range commensurate to Figure \ref{HIST_dist_reach} namely ($100$ \Mpc - $1$ Gpc). 

\section{Multi-detector sky coverage, maximum distance reach and the fractional detection volume}\label{SC}

In this section, we compare the observing ability of all the possible multi-detector networks. We define figures of merit such as the sky coverage,  the maximum distance reach ($R_{h}$) and  the fractional detection volume  ${\cal V}$ for each network \cite{Schutz:2011tw}. All these quantities depend on the binary inclination. Here we assume that all the binaries are distributed uniformly in $\epsilon$ and $\psi$. Then for a network of $I$ detectors, the multi-detector SNR square averaged over $\epsilon$ and $\psi$ gives, 

\be
\overline{\Ro^2_{\epsilon,\Psi}} = \frac{14}{15} \sum_{m=1}^{I} \f_{+m}^2 + \frac{2}{3} \sum_{m=1}^{I} \f_{\times m}^2 \sim \sum_{m=1}^{I} \f_{+m}^2 + \sum_{m=1}^{I} \f_{\times m}^2 \,. \label{snr}
\ee
This is very similar to the face-on case ($\epsilon = 0$). Hence, below we take the face-on case.

{\it Maximum distance reach:} The maximum distance reach ($R_h$) is the maximum distance observed by a network in its most sensitive direction for a given source. This is determined by the choice of threshold SNR $\Ro_{th}$. If $R(\theta,\phi,\Ro_{th})$ is the distance reach of a network in the given direction,  then $R_h$ is the  maximum of $R(\theta,\phi, \Ro_{th})$.

{\it Sky Coverage:} The isotropy in directional response of a multi-detector network can be quantified  in terms of sky coverage -- the percentage of  sky area {\it visible} to a multi-detector network. 
We define the sky coverage as the fraction of sky area with the network SNR $\Ro$ greater than the threshold value $\Ro_{th}$.

{\it Fractional sky volume ${\cal V}$:} For a given network with the threshold network SNR of $\Ro_{th}$,
we define
\be
{\cal V} = \int_0^{R(\theta,\phi, \Ro_{th})} \int_0^\pi \int_0^{2 \pi} r^2 \sin \theta~dr~ d \theta~ d \phi  \,.
\ee

The fractional detection
volume ${\cal V}$, following \cite{Schutz:2011tw},  measures the 3 dimensional detectable sky volume of a given network. 

The Table \ref{Chap3.SkyCoverage} gives quantitative measurement of the sky coverage (Column III), maximum distance reach (Column IV) and fractional detection volume of all possible 2 - 5 detector combinations for non-spinning NS-BH binaries. For computing the maximum distance reach and  sky coverage, we take the threshold SNR $\Ro_{th} = 6$ which is also represented graphically in the Figure  \ref{Chap3.SkyCoverage2}.

To explain the trends in table \ref{Chap3.SkyCoverage}, we consider a pair of  detectors in the given network. We define the angle between the two detector (i and j) planes in that pair (or local zenith) as $\delta_{ij}$. This angle is tabulated for different
detector pairs in Table \ref{Chap3.SkyCoverage} (Column II in bracket). 
Using this we define a geometrical quantity 
\be
{\mathcal G} \equiv \sqrt{I} [1 + (I-1) <\cos \delta_{ij} >]^{1/2} \, ,
\ee

For an $I$ detector network, $\sqrt{I} \le {\mathcal G} \le I$. The $<\cdot>$ denotes average over all possible ${}^I C_2$ pairs of detectors in that network. 
Please note, when all the detectors in a network are
aligned ($<\cos \delta_{ij}>=1$), ${\mathcal G} = I$. 

For aligned detectors (the detectors are oriented in a particular direction) the network has  better reach in that sensitive direction giving large $R_h$. However, in other directions, the network is not very sensitive making the overall sky coverage low. 
We explain the sky coverage and $R_h$ for different networks using ${\mathcal G}$ computed for each network tabulated in
Table \ref{Chap3.SkyCoverage} (Column II).
In Table \ref{Chap3.SkyCoverage} (Column V), we tabulate the fractional detection volume for 
each network as compared with the single LIGO-L detector. The trends in fractional detection volume are  similar to the maximum
distance reach $R_h$. This is expected because higher is the distance reach, larger is the detection volume expected to
cover.

\subsection{Two detectors} For two detector networks, the ${\mathcal G}$ ($= \sqrt{2 (1 + cos \delta)}\  $)  depends on the angle between the detector planes. Thus, the sky-coverage increases with increase in the $\delta$  (See the trend in sky coverage). Further, we see the decrease in $R_h$  as the angle between the two planes increases. This is natural, as for the nearly aligned detectors ({\it i.e.} low $\delta$, e.g. LH), the distance reach would be highest along the most sensitive direction. However, the $\delta$ for VK pair is maximum, in fact the V and K planes are almost orthogonal ($\delta=86.62$). This in turn gives close to the least maximum distance reach for VK pair amongst all the 2 detector networks. Please note that LK and HV show slightly higher sky coverage compared to VK in spite of lower $\delta$. This can be attributed to the in-plane orientation of the detectors. In summary;\textit{ amongst the 2 detector networks HV (LH) gives maximum (minimum) sky coverage and 
	LH (VK) gives the maximum (minimum)  distance reach} [Figure (\ref{Chap3.SkyCoverage2}.a)].

\subsection{Three detectors} For a 3 detector network, $\sqrt{3} \le {\mathcal G} \le 3$. The $\mathcal G=\sqrt{3}$ case corresponds to the detectors with antenna-patterns forming an orthogonal triplet.
We note that HVK and LVK  combinations give maximum sky coverage ($100\%$) and minimum $R_h$ namely $1.53 \ \Gpc$ and $1.49\ \Gpc$,  which attributes to ${\mathcal G}$ closer to the lower end of the spectrum making them close to orthogonal triplets. The $\delta_{ij}$ numbers between different pairs in HVK and LVK network supports the same. The combination LHI gives minimum sky coverage ($84.55\%$) and maximum $R_h$ $(1.84 \ \Gpc)$, which attributes to the higher value of ${\mathcal G}$ (close to 3). 
\textit{In the 3 detector networks HVK and LVK (LHI) gives maximum (minimum) sky coverage and LHI (LVK) gives the maximum (minimum) maximum distance reach} [Figure (\ref{Chap3.SkyCoverage2}.a)].

\subsection{Four detectors} 
\begin{figure*}[ht]
	\centering
	\includegraphics[width=\textwidth]{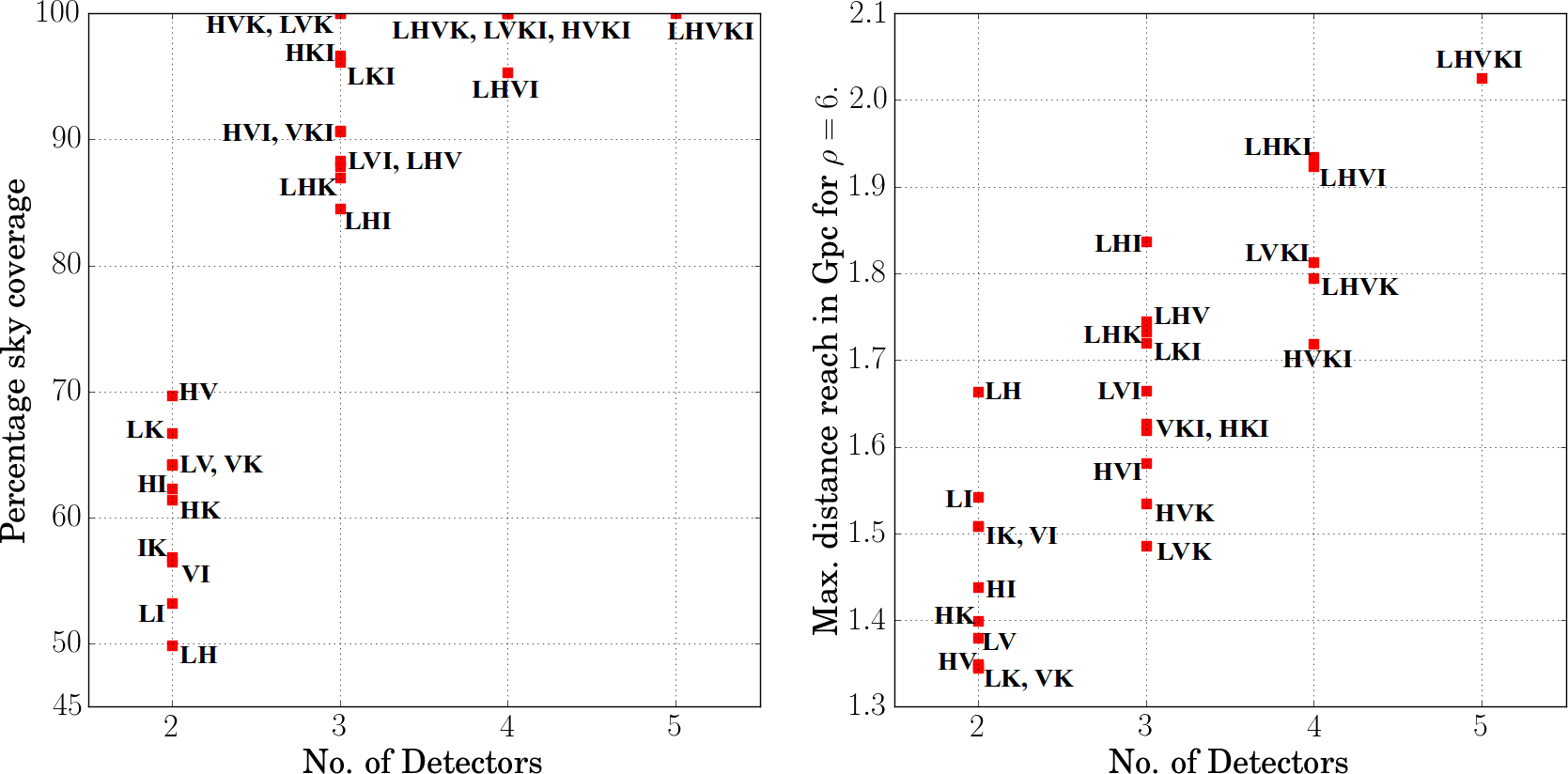}
	\caption{ \label{Chap3.SkyCoverage2} (a) Sky coverage of various networks. (b) Maximum distance reach of various networks for ~$(1.4-10)M_\odot$ NS-BH systems with $\epsilon = 0$ and $\Psi = 0$.}
\end{figure*}

The networks LHVK, LVKI, LHKI and HVKI show $100\%$ sky coverage. 
LHVI shows slightly low sky coverage, $95.34\%$.  The trend in the distance reach is reverse to that of the sky coverage. \textit{Amongst the 4 detector networks LHVK, LVKI, LHKI and HVKI (LHVI) gives maximum (minimum) sky coverage and LHVI (LHVK) gives the maximum (minimum) maximum distance reach} [Figure (\ref{Chap3.SkyCoverage2}.a)].

\subsection{Five detectors} The 5 detector gives the maximum distance reach of $2.03 \ \Gpc$ almost $1.22$ times deeper than the two detector LH and $1.51$ times deeper than LK.  LHVKI covers sky volume $3.6$ times that of LH and $3.8$ times that of LK. This would translate into the injection recovery with these networks which
we will discuss in the following section.

For NS-NS system,  the distance reach would scale by a factor $\sim \mathcal{M}_c^{5/6}$ and the absolute detection volume would scale as $\sim \mathcal{M}_c^{5/2}$, where $\mathcal{M}_c$ is the {\it chirp mass} of the system. For an NS-NS system ($\mathcal{M}_c=1.22M_{\odot}$) distance reach will be scaled by a factor $\sim 0.473$ and detection volume will be scaled by a factor $\sim 0.106$ . Here, we assume that the noise PSDs in each detector are identical. However, with real advanced detector noise curves, there would be small variations in the above estimates.


\begin{table}[h]
\centering
		\begin{tabular}{l c c c c}
			\hline
			Network & ${\mathcal{G}} (\delta~in~deg.)$ & Sky Coverage & $R_{h}$ in Gpc & ${\cal V}/{\cal V}_L$ \\ \hline
			LH  & 1.94 (27.25) & 49.85\% & 1.66 & 2.77\\ 
			LI  & 1.8 (51.76)  & 53.23\% & 1.54 & 2.65\\ 
			VI  & 1.75 (57.99) & 56.52\% & 1.51 & 2.63\\ 
			KI  & 1.75 (57.46) & 56.87\% & 1.51 & 2.62\\ 
			HK  & 1.61 (72.56) & 61.42\% & 1.4 & 2.57\\ 
			HI  & 1.67 (66.67) & 62.33\% & 1.44 & 2.59\\ 
			LV  & 1.57 (76.76) & 64.21\%  & 1.38 & 2.57\\ 
			VK  & 1.46 (86.62) & 64.23\% & 1.35 & 2.55\\ 
			LK  & 1.52 (80.56) & 66.74\%  & 1.34 & 2.56\\ 
			HV  & 1.54 (79.63) & 69.72\% & 1.35 & 2.56\\ 
			\hline
			LHI & 2.61 & 84.55\% & 1.84 & 4.81\\ 
			LHK & 2.39 & 86.97\% & 1.73 & 4.73\\ 
			LVI & 2.40 & 87.86\% & 1.66 & 4.67\\ 
			LHV & 2.36 & 88.28\% & 1.75 & 4.72\\ 
			HVI & 2.28 & 90.66\% & 1.58 & 4.62\\ 
			VKI & 2.30 & 90.71\% & 1.62 & 4.65\\ 
			LKI & 2.37 & 96.14\% & 1.72 & 4.65\\ 
			HKI & 2.34 & 96.65\% & 1.63 & 4.63\\ 
			HVK & 2.02 & 100.0\% & 1.53 & 4.53\\ 
			LVK & 1.97 & 100.0\% & 1.49 & 4.51\\ 
			\hline
			LHVI & 3.07 & 95.34\% & 1.92 & 7.14\\ 
			LHKI & 3.12 & 100.0\% & 1.93 & 7.16\\ 
			LVKI & 2.88 & 100.0\% & 1.81 & 7.01\\ 
			HVKI & 2.82 & 100.0\% & 1.72 & 6.96\\ 
			LHVK & 2.76 & 100.0\% & 1.8 & 7.0\\ 
			\hline
			LHVKI  & 3.58 & 100\%  & 2.03 & 9.77\\ \hline 
		\end{tabular}
	\caption{${\cal G}$, Sky Coverage, Maximum distance reach ($R_h$) in Gpc and the fractional detection volume with respect to single detector L (${\cal V}/{\cal V}_L$) for all possible 2-5 detector network configurations. We choose the $\Ro_{th} =6$ for NS-BH system with masses $(1.4-10)M_\odot$ located at $1 \ \Gpc$.}
	\label{Chap3.SkyCoverage}
\end{table}

\section{Coherent multi-detector CBC search}\label{statistics}
The multi-detector searches for compact binaries are generally carried out in two distinct ways;  coincidence approach  and coherent approach.  In coincidence scheme, data from each interferometer is processed individually and candidate events are listed by comparing individual SNRs  with the threshold. Then the  recorded events in individual detectors are compared for coincidences in  mass and time of arrival parameters \cite{ihope,Abadie2012}.  On the other hand, in coherent search strategy, the data from different detectors  are combined in a phase coherent fashion into a single effective network statistic and a detection would be carried out by applying a threshold on it  \cite{Pai:2000zt,Macleod2016, Harry:2010fr}. In other words, coherent search scheme combines the GW signal power from individual detectors  to form  effective  multi-detector network SNR.  In the literature, it has been demonstrated that the coherent search performs better than the coincidence search for coalescing  binaries  \cite{tagoshi2006}. The simulations show that the coherent scheme gives  $\sim 30 \%$ improvement in sensitivity compared to the coincidence approach \cite{Harry:2010fr}. 

Currently the targeted coherent search is implemented as a follow up analysis for the  IPN-GRB search \cite{PhysRevLett.113.011102}. In \cite{Haris:2015iin} authors proposed a new multi-detector semi-coherent statistic; \textit{hybrid} statistic as an alternative for the well established MLR statisticc.
Below we summarize the fully-coherent MLR statistic and the semi-coherent \textit{hybrid} statistic. 

\subsection{Maximum likelihood ratio statistic (MLR):}
The multi-detector MLR statistic is  constructed by explicitly  maximizing the network log likelihood ratio statistic over four extrisic CBC signal parameters namely, constant amplitude, binary inclination angle $\epsilon$, polarization angle $\Psi$ and the initial phase of the signal \cite{Pai:2000zt,Harry:2010fr,Haris:2014fxa}.  Exploiting the orthogonality property of antenna pattern functions in the dominant polarization frame, we can express the multi-detector MLR statistic in terms of two synthetic streams $\z_L$ and $\z_R$. The synthetic streams are constructed by linearly combinig 
data from different detectors as below.

\be
\tilde {\tilde z}_L(f) \equiv \sum_{m=1}^I \frac{\f_{+ m}}{\|\F^{'}\|} \tilde{\tilde x}_m(f),~~\tilde{\tilde z}_R(f) \equiv \sum_{m=1}^I \frac{\f_{\times m}}{\|\F^{'}\|} \tilde{\tilde x}_m(f)\, .  \label{zLzR}
\ee 
Here $\tilde{\tilde x}_m(f)$ denotes  the frequency domain over-whitened data from $m$-th detector.  In terms of $\z_L$ and $\z_R$, the MLR statistic is given by
\be
\Lc  =  \langle  \z_L |  \h_0 \rangle^2 + \langle  \z_L |  \h_{\pi/2} \rangle^2 + \langle  \z_R |  \h_0 \rangle^2 + \langle  \z_R |  \h_{\pi/2} \rangle^2 \,.
\label{MLR1} 
\ee
The $\f_m \equiv \f_{+m} +i \f_{\times m} $ is the complex antenna pattern function of the $m$-th detector (in the dominant polarization frame) \footnote{For details about the dominant polarization frame, please refer to \cite{Haris:2014fxa}}. The quantity {\small $\|\F^{'}\|^2 \equiv {\sum_{m} \langle \h_0|\h_0\rangle_m~ (\f_{+m}^{2} + \f_{\times m}^{2})} $} is the noise weighted norm  of the complex antenna pattern vector of the network of $I$ detectors.  Physically, it captures the effective GW power transfered to the interferometric network averaged over $\epsilon$ and $\psi$. The statistic follows $\chi^2$ distribution with 4 degrees of freedom in the absence of signal. The MLR statistic captures the optimum multi-detector SNR $\Ro$ of the signal expressed in Equation \ref{signal}. However, the presence of noise in the two data streams constructed  in the formalism increases the false alarms. This will increase the false alarm probability (FAP) of the statistic \cite{Pai:2000zt,Haris:2014fxa}.

\subsection{Hybrid statistic:}
In \cite{Haris:2015iin} authors introduced a semi-coherent statistic known as the \textit{hybrid} statistic. The \textit{hybrid} statistic is defined as the maximum of  the multi-detector MLR statistics specially tuned for face-on ($\epsilon=0$) and face-off ($\epsilon = \pi$) binary systems and it can be expressed in terms of a single synthetic data stream. This will reduce the FAP. As it captures less SNR, the statistic is semi-coherent. For a wide range of $\epsilon$ it can capture  optimum multi-detector SNR $\Ro$ of the signal. The \textit{hybrid} statistic $\Lc^{mx}$ is defined as 
\be
\Lc^{mx} \equiv \max\{\Lc^0,\Lc^\pi\}~, \label{hyb}
\ee
where $\Lc^0(\Lc^{\pi})$ is the MLR statistic for face-on(off) binary is given by,
\be
\Lc^{0,\pi} = \langle  \z^{0,\pi} |  \h_0 \rangle^2 + \langle  \z^{0,\pi} |  \h_{\pi/2} \rangle^2~. 
\ee
This statistic follows $\chi^2$ distribution with 2 degrees of freedom in the absence of the signal. Here the over whitened synthetic data streams are,
{\small \be
	\tilde {\tilde z}^0(f)  \equiv \sum_{m} \frac{\f_m}{\|\F^{'}\|} \tilde{\tilde x}_m(f),~~~ 
	\tilde {\tilde z}^\pi(f)  \equiv \sum_{m} \frac{\f^{*}_m}{\|\F^{'}\|} \tilde{\tilde x}_m(f) . \label{Z0Zpi}  
	\ee}

In \cite{Haris:2015iin}, the authors show with Gaussian noise in LHV network that,  the single stream {\it hybrid} statistic gives less \FAP compared to the MLR statistic with little compromise on the detection probability (DP). The Receiver Operator Characteristic (ROC) curves clearly show that for a wide range of binary inclination angles $(\epsilon<70^\circ ~\text{and}~\epsilon>110^\circ)$ the {\it hybrid} statistic gives better performance compared to the MLR statistic .

In  \cite{Williamson:2014wma}, a similar statistic (similar to Equation \ref{hyb}) was used for the  GW follow-up of short Gamma Ray Bursts (sGRBs) of IPN triggers. This was targeted search with templates in mass parameter space in LIGO-Virgo data. The study shows a similar improvement in the false alarm rates compared to generic MLR statistic on heuristic grounds. The sGRBs are expected to have narrow opening angle $(<30^\circ)$. Therefore sGRBs will be visible to gamma ray telescopes only when the line of sight nearly coincide with the axis of GRB. This is the reason why in \cite{Williamson:2014wma}, the authors used MLR tuned for face-on/off binaries for the search of IPN-GRBs.  However, in \cite{Haris:2015iin} authors have shown that {\it hybrid} statistic can preform better than the generic MLR statistic for much wider region of inclination angle and hence it can be used for the  search of binary inspirals having arbitrary inclination angle without compromising too much on the SNR.

\section{Performance comparison  between $\Lc$ and $\Lc^{mx}$}\label{simulations}

\begin{figure*}
	\centering
	\includegraphics[width=0.7\textwidth]{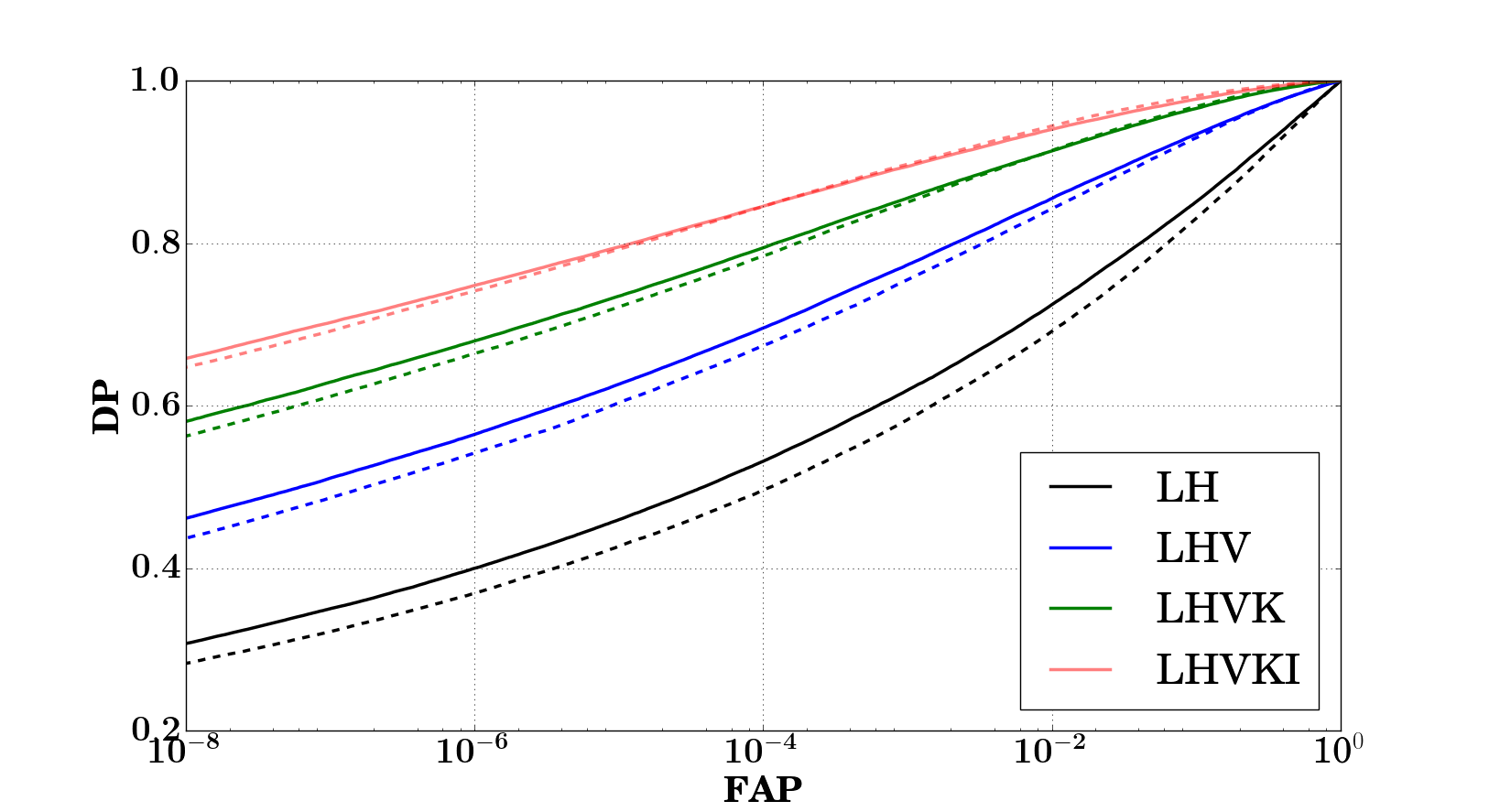}
	\caption{ \label{Chap4.MLR_ROC} ROC curves of $\Lc$ (dashed curve) and $\Lc^{mx}$ (solid curve) for networks LH, LHV, LHVK and LHVKI. The DNS binary injections of masses $(1.4 - 1.4)~M_\odot$ are distributed uniformly in the sky volume in a distance range $(100 \ \Mpc-1 \ \Gpc)$. The binary orientation $\cos \epsilon$  and $\Psi$ are uniformly sampled from the ranges $(-1,1)$ and  $(0, 360^\circ)$ respectively. All detectors are assumed with \enquote{zero-detuning, high power} Advanced LIGO PSD noise curve  \cite{aLIGOSensitivity}.}
\end{figure*}

In this section, we carry out numerical simulations to estimate the detection performance of  $\Lc^{mx}$ and $\Lc$ for all combinations of 2, 3, 4 and 5 detector networks composed of L, H, V, K, and I. We perform the simulations for three kinds of binary systems;\\

\noindent \textbf{DNS}: Non-spinning neutron star binaries with masses $(1.4-1.4) M_{\odot}$.\\

\noindent \textbf{NS-BH}: NS-BH binary systems of masses $(1.4-10)M_{\odot}$ with non-spinning as well as spinning BH and non-spinning NS component. The spin of the BH component  is  sampled uniformly from the range $(-1,1)$ for spinning systems. \\

For both DNS and NS-BH injections,  the inclination angle $\epsilon$ and polarization angle $\psi$ are sampled uniformly from the polarization sphere.  The sources are distributed uniformly in sky volume within a distance range $(100 \ \Mpc - 1 \ \Gpc)$. 
The total number of sample points is $10^5$ for all the cases and is denoted by sample-1. The subset of them belong to  $(\epsilon \leq 70^\circ~or~ \epsilon \geq 110^\circ)$  is denoted by sample-2.

All the detectors are assumed to have Gaussian, random noise with the noise PSD following \enquote{zero-detuning, high power} Advanced LIGO noise curve \cite{aLIGOSensitivity}. However, since the  distribution of both $\Lc$ and $\Lc^{mx}$ depends on the only noise model, here we use the analytic expressions to compute the  FAP of $\Lc$ and $\Lc^{mx}$ for Gaussian noise model. Following  Equation.10 and Equation.33 of \cite{Haris:2015iin}, the probability distributions of $\Lc$ and $\Lc^{mx}$ in the absence of the signal are given by,
\bea
\text{P}_{\Lc}(\pounds) &=& \frac{\pounds}{4} \exp[-\pounds/2]~, \nn \\
\text{P}_{\Lc^{mx}}(\pounds) &=&  \frac{1}{2~g}~  e^{-\frac{\pounds^{mx} }{2 \ll( 1- g^2 \rr)}} \int_0^{\pounds^{mx}}  e^ {-\frac{\pounds^\pi }{2 \ll( 1- g^2 \rr)}} ~I_0 \ll(\frac{g}{1-g^2} \sqrt{\pounds^{mx} \pounds^\pi}\rr)~ d\pounds^\pi,
\eea
where $g$ is a constant. The FAPs can be computed by integrating above equations. The corresponding   DP is computed numerically by counting the number of times the statistic crosses a given threshold.  

Figure \ref{Chap4.MLR_ROC} shows the ROC curves of $\Lc$ and $\Lc^{mx}$ for LH, LHV, LHVK and LHVKI detector networks which act as representative 2, 3, 4 and 5 detector networks for DNS system. We clearly see that the ROC curves for $\Lc^{mx}$ (solid curve) are above that of the $\Lc$ (dashed curve). This means that for a given \FAP, the \DP  of $\Lc$ is higher than that of $\Lc$ indicating better injection recovery.

\begin{table}[h]
	\begin{center}
		\begin{tabular}{l c c c c}
			\hline
			\multirow{2}{*}{Network} & \multicolumn{2}{c}{$\%$ of Injections recovered in} & \multicolumn{2}{c}{Avg. distance reach in Mpc for} \\ 
			& ~~$\Lc$~~ & ~~$\Lc^{mx}$~~~~&~~~~~$\Lc$~~~~~ & ~~$\Lc^{mx}$~~~~\\ \hline
			LH  &  ~~5.909~~ & 6.893  &   435 &	458\\
			LI  &  5.707 & 6.452 & 410 	& 434\\
			KI  &  5.760 & 6.549 & 406 	& 432\\
			VI  &  5.677 & 6.515 & 409 	& 433\\
			HI  &  5.660 & 6.418 & 399 	& 420\\
			HK  &  5.407 & 6.290 & 403 	& 421\\
			LV  &  5.310 & 6.046 & 399 	& 419\\
			VK  &  5.389 & 6.234 & 392 	& 416\\
			LK  &  5.476 & 6.239 & 392 	& 415\\
			HV  &  5.5765 & 6.411 & 388 	& 411\\            
			\hline
			LHI &  10.18 & 11.49 & 489 	& 519\\
			LKI &  9.786 & 11.07 & 475 	& 498\\
			LHV &  9.836 & 11.16 & 479 	& 507\\
			LHK &  9.931 & 11.32 & 480 	& 505\\
			LVI &  9.859 & 11.14 & 470 & 493\\
			HKI &  9.948 & 11.29 & 463 	& 489\\
			VKI &  9.851 & 11.12 & 472 	& 494\\
			HVI &  10.08 & 11.27 & 472 	& 502\\
			HVK &  9.837 & 11.02 & 450 	& 473\\
			LVK &  9.865 & 11.10 & 449 	& 476\\
			\hline
			LHKI &  14.73 & 16.37 & 536 & 	563\\
			LHVI &  15.03 & 16.82 & 531 & 	561\\
			LHVK &  14.46 & 16.20 & 529 & 	554\\
			LVKI &  14.57 & 16.16 & 521 & 	549\\
			HVKI &  14.84 & 16.55 & 519 & 	545 \\
			\hline            
			LHVKI&  19.86 & 21.90 (30.88)  & 577 & 	608
			\\ \hline  
		\end{tabular}
	\end{center}
	\caption{ Percentage of total injections recovered by $\Lc$ and $\Lc^{mx}$ and their average distance reach for various networks with \FAP$={10^{-7}}$. The DNS binary injections of masses $(1.4 - 1.4)~M_\odot$ are distributed uniformly in the sky volume in the distance range $(100 \ \Mpc-1 \ \Gpc)$. $\cos \epsilon$  and $\Psi$ are uniformly sampled from the ranges $(-1,1)$ and  $(0, 360^\circ)$ respectively. All detectors are assumed with \enquote{zero-detuning, high power} Advanced LIGO PSD  noise curve \cite{aLIGOSensitivity}.}
	\label{Chap4.DetectionRate5}
\end{table}

\begin{table}
\begin{center}
	\begin{tabular}{lcccc}
		\hline
		\multirow{3}{*}{Network} & \multicolumn{4}{c}{\% of injection recovered}\\
		& \multicolumn{2}{c@{\quad}}{Spinning NS-BH}    
		&                                          
		\multicolumn{2}{c}{Non-spinning NS-BH} \\
		&$\Lc$ & $\Lc^{mx}$    & $\Lc$ & $\Lc^{mx}$      \\
		\hline
		LH  &  32.57  & 35.27 &  32.74 & 35.50  \\
		LI  &  33.43  & 35.60 &  33.85 & 36.07  \\
		KI  &  33.82  & 36.32 &  34.28 & 36.81  \\
		VI  &  33.59  & 36.22 &  33.72 & 36.29  \\
		HI  &  34.32  & 36.81 &  34.42 & 37.04  \\
		HK  &  34.52  & 37.28 &  34.61 & 37.31  \\
		LV  &  34.47  & 36.40 &  34.72 & 37.39  \\
		VK  &  34.73  & 37.50 &  35.17 & 37.87  \\
		LK  &  34.42  & 37.08 &  34.91 & 37.58  \\
		HV  &  34.68  & 37.11 &  35.13 & 37.62  \\            
		\hline
		LHI &  47.76  & 50.02 &  47.94 & 50.06  \\
		LHK &  48.88  & 51.34 &  49.05 & 51.55  \\
		LVI &  49.25  & 51.35 &  49.46 & 51.51  \\
		LHV &  48.89  & 51.22 &  48.87 & 51.39  \\		
		HVI &  49.89  & 52.06 &  50.12 & 52.35  \\			
		VKI &  49.48  & 52.02 &  49.93 & 52.52  \\		
		LKI &  49.27  & 51.38 &  49.81 & 51.98  \\				
		HKI &  49.52  & 51.88 &  50.34 & 52.75  \\
		HVK &  51.01  & 53.34 &  51.36 & 53.72  \\
		LVK &  51.23  & 53.54 &  51.74 & 54.01  \\		
		\hline
		LHVI &  59.99 & 61.51 &  60.35 & 61.91  \\		
		LHKI &  60.37 & 61.99 &  60.70 & 62.40  \\		
		LVKI &  61.22 & 62.77 &  61.99 & 63.52  \\
		HVKI &  61.69 & 63.42 &  62.05 & 63.62  \\
		LHVK &  61.87 & 63.56 &  61.95 & 63.67  \\
		\hline            
		LHVKI&  69.39 & 70.27 &  69.96 & 70.84  \\
		\hline  
	\end{tabular}
\end{center}
	\caption{ Percentage of total injections recovered by $\Lc$ and $\Lc^{mx}$ for various networks with \FAP$={10^{-7}}$. Columns II and III are for  aligned spin NS-BH binary injections  and columns  IV and V are for non-spinning NS-BH binary injections of masses $(1.4 - 10)~M_\odot$ are distributed uniformly in the sky volume in the distance range $(100 \ \Mpc-1\ \Gpc)$. $\cos \epsilon$  and $\Psi$ are uniformly sampled from the ranges $(-1,1)$ and  $(0, 360^\circ)$ respectively. Spin of the BH component is uniformly sampled from the range $(-1,1)$. All detectors are assumed with \enquote{zero-detuning, high power} Advanced LIGO PSD  noise curve \cite{aLIGOSensitivity}.}
	\label{Chap4.DetectionRate5NS_BH}
\end{table}

In Table \ref{Chap4.DetectionRate5} and \ref{Chap4.DetectionRate5NS_BH}, we make  quantitative comparison of the performances of $\Lc$ and $\Lc^{mx}$ for various networks. We choose
the  FAP $=10^{-7}$ for computing the injection recovery. This corresponds to the SNR threshold of $6.18$ for $\Lc$ and  SNR threshold of 5.79 for $\Lc^{mx}$. In the second and third columns respectively, we list the percentage of injections recovered  by $\Lc$ and $\Lc^{mx}$. Below we summarize the salient features of the simulation result.

The Table \ref{Chap4.DetectionRate5} contains the percentage of recovered DNS injections and  the Table \ref{Chap4.DetectionRate5NS_BH} contains the percentage recovery  of  spinning as well as  non-spinning NS-BH injections. The first observation one can make from the tables is that  the percentage injection recovery for both statistics are higher for NS-BH injections compared to the DNS injections. This is because of the fact that for NS-BH injections the total mass of the system is higher than that of DNS system. Therefore if all other parameters are fixed,  the NS-BH injections  gives higher SNR in the network. The relative increase in SNR of NS-BH systems compared to DNS systems results in  higher detection probability (or higher injection recovery).

The Columns IV and V of  Table \ref{Chap4.DetectionRate5} gives the average luminosity distance of injections recovered by $\Lc$ and $\Lc^{mx}$ for various networks with the DNS injections. For all networks, the average luminosity distance for  $\Lc^{mx}$ is higher than that of $\Lc$. This is due to the fact that the sensitivity of $\Lc^{mx}$ is higher than that of $\Lc$. Further, as expected, the number roughly follows the trend of $R_{h}$ tabulated in   Table \ref{Chap3.SkyCoverage}. For MLR statistic the average luminosity distance for 2, 3, 4,and 5 detectors networks are 403 Mpc, 469 Mpc, 554 Mpc, and 608 Mpc respectively. On the other hand, for $\Lc^{mx}$ the numbers are 425 Mpc, 495 Mpc, 554 Mpc, and 608 Mpc respectively. 

\subsection{Variation in injection recovery with number of detectors in the network}
We  note that the injection recovery (Columns  II and III in Table \ref{Chap4.DetectionRate5} and Columns II-V in Table \ref{Chap4.DetectionRate5NS_BH}) increases with the number of detectors in the network for both $\Lc$ and $\Lc^{mx}$ (see  Figure \ref{Chap4.MLR_ROC}). For the non-spinning DNS injections, the typical average\footnote{Average is taken over all possible detector combinations with a fixed number of detectors} injection recovery using $\Lc$ ($\Lc^{mx}$) are $\sim 5.587\%$ ($6.404\%$),  $9.917$ ($11.20\%$), $14.73\%$ ($16.42\%$) and $19.86\%$ ($21.90\%$)  for 2, 3, 4 and 5 detector networks respectively. 
For spinning NS-BH injections the corresponding numbers are given by  $\sim 34.55\%$ ($36.56\%$),  $49.52\%$ ($51.81\%$), $61.09\%$ ($62.65\%$) and $69.39\%$ ($70.27\%$).
For non-spinning NS-BH injections  the numbers shows a slight deviation from  the spinning injections and are given by $\sim 34.35\%$ ($36.95\%$),  $49.86\%$ ($52.18\%$), $61.41\%$ ($63.02\%$) and $69.96\%$ ($70.84\%$) for 2, 3, 4 and 5 detector networks respectively.
The increase in injection recovery with number of detectors is  primarily due to the fact that  the network SNR $\Ro$ increases with the number of detectors, which in turn increases the distance reach as well as the detection volume of the network. The successive addition of detectors one-by-one to a typical 2-detector network increases the injection recovery for spinning NS-BH system by $\sim 50.11\%(45.22\%)$ for 3 detectors, $\sim 89.96\%(80.21\%)$ for 4  detectors and $ \sim 103.84\%(99.23\%)$ for 5 detector network with respect to the 2 detector network.
\subsection{Variation in injection recovery with detector combinations}
For networks with a fixed number of detectors, the injection recovery (for both $\Lc$ and $\Lc^{mx}$) increases with the sky coverage  of the network listed in Table \ref{Chap3.SkyCoverage} (Column III). 
Among 3 detector networks, LVK and HVK show higher values of injection
recovery (due to their high sky coverage). Similarly HVKI shows high recovery amongst the 4 detector networks.

\subsection{Improvement of injection recovery with $\Lc^{mx}$ over $\Lc$}

For all sets of injections, we observe that the $\Lc^{mx}$ recovers more injections compared to $\Lc$. However, the improvement in injection recovery of $\Lc^{mx}$ decreases with increase in the number of detectors in the network. 
The average fractional improvement in the injection recovery of $\Lc^{mx}$ over $\Lc$  for DNS injections are  $\sim14.63\%$, $\sim 12.91\%$, $\sim 11.49\%$ and $\sim 10.29\%$ for 2, 3, 4 and 5 detector networks respectively.
The corresponding numbers for spinning (non-spinning) NS-BH injections  are given by  $\sim 7.35\%$ ($\sim 7.55\%$), $\sim 4.48\%$ ($\sim 4.66\%$), $\sim 2.5\%$ ($\sim 2.63\%$) and $\sim 1.19\%$ ($\sim 1.26\%$) for 2, 3 ,4 and 5 detector networks respectively.

As discussed earlier,
as the number of detectors increases, the SNR increases and hence the injection recovery  for both $\Lc$ and $\Lc^{mx}$ statistics increases. However, as we note earlier, the performance improvement of $\Lc^{mx}$ over $\Lc$ is through the reduced FAP, which is independent of SNR. Therefore the advantage of $\Lc^{mx}$ over $\Lc$ reduces with number of detectors in the network.

\subsection{ Injection recovery with binary inclination  and distance }

\begin{figure}[ht]
	\centering  	
	\includegraphics[width=0.8\textwidth]{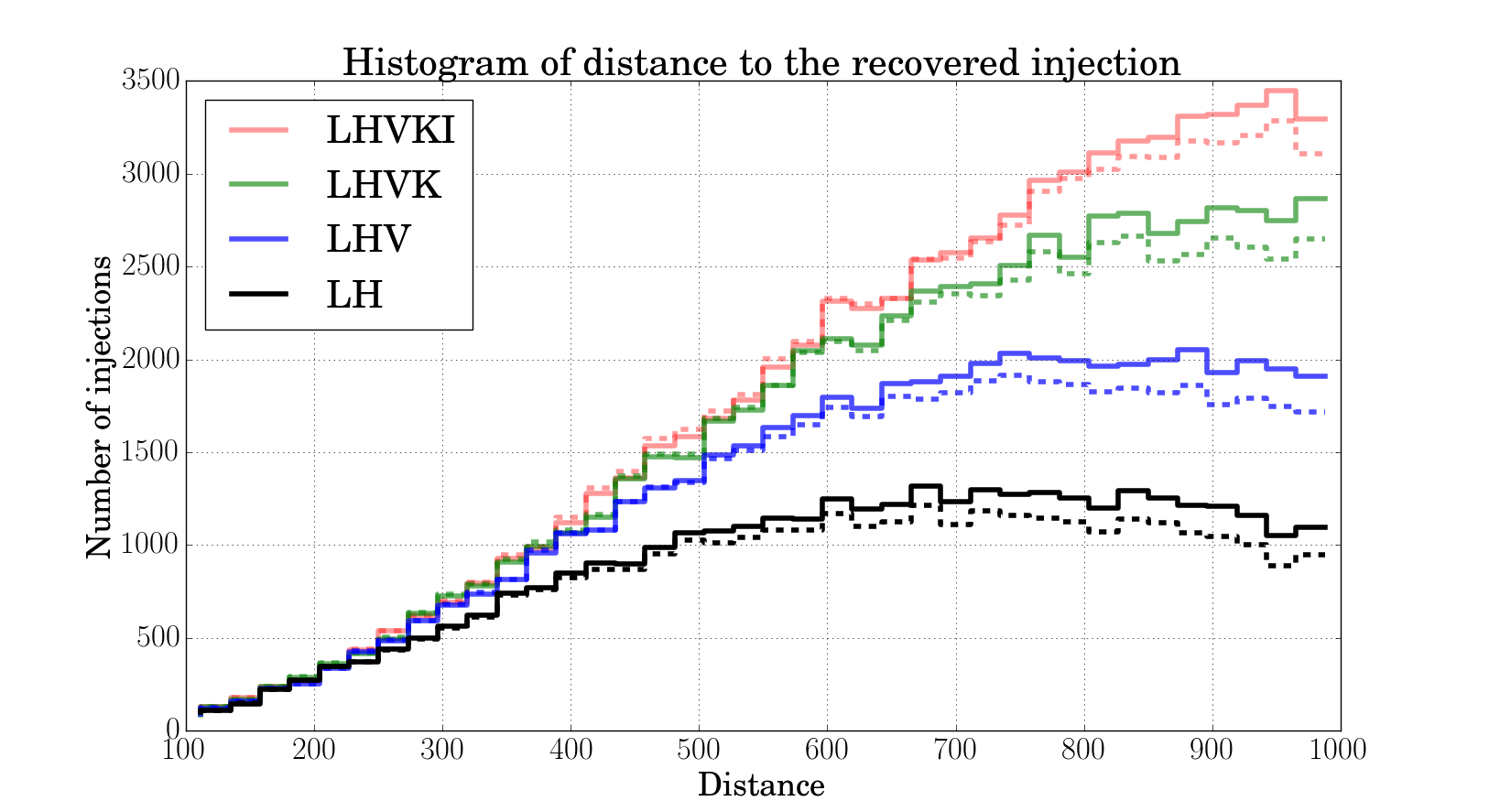}
	\caption{Distribution of distance of injections recovered using  statistic $\Lc$ (dashed line) and $\Lc^{mx}$ (solid line) for different detector networks.} \label{Chap4.Rec_D_Hist}
\end{figure}

\begin{figure}[ht]
	\centering  	\includegraphics[width=\textwidth]{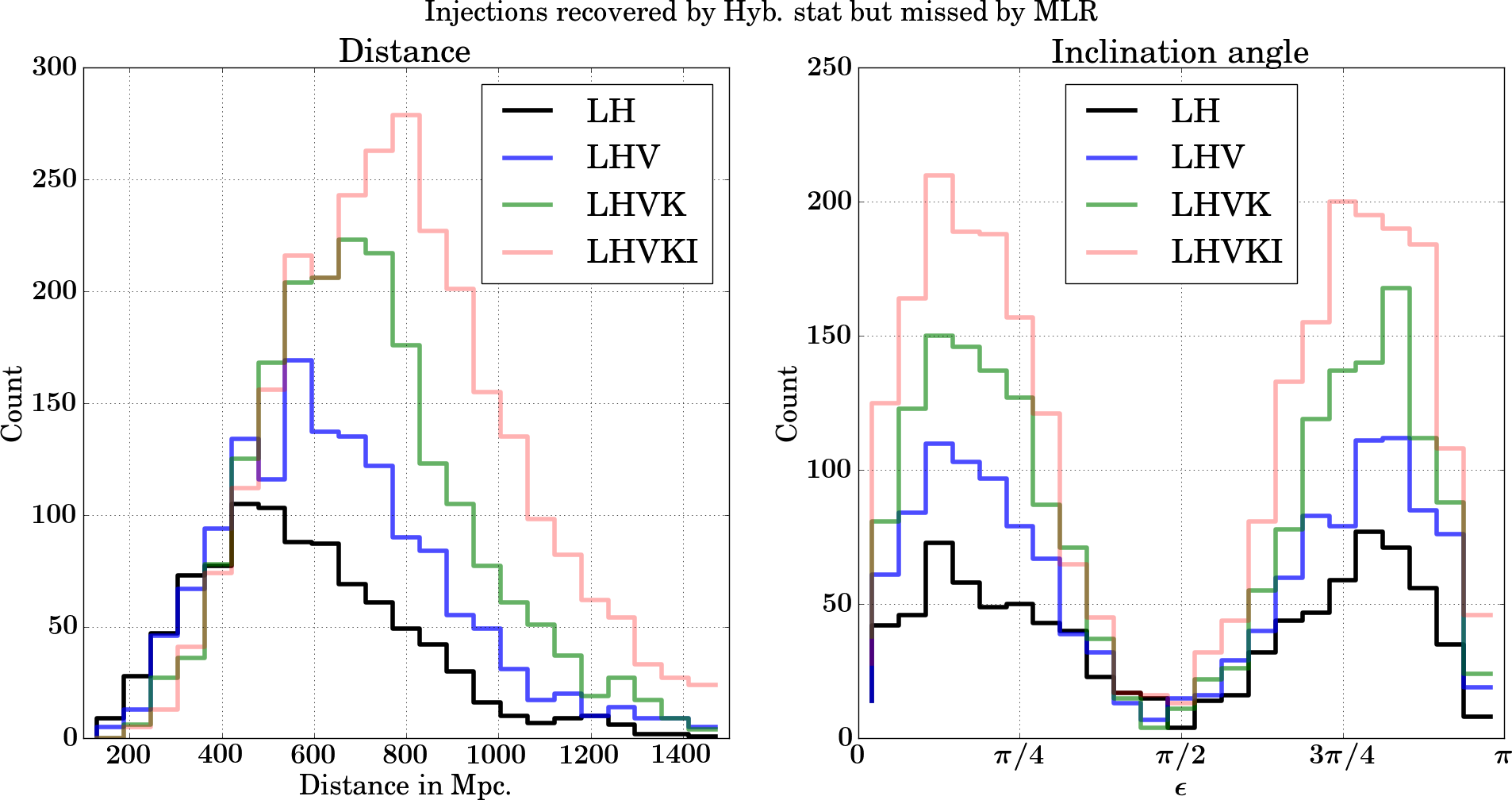}
	\caption{ Histograms of distance and inclination angle of  injections recovered by $\Lc^{mx}$, but missed by $\Lc^{mx}$ for various networks.} \label{Chap4.Drate_Epsi_2}
\end{figure}	

Figure \ref{Chap4.Rec_D_Hist} shows the histogram of luminosity distance of the  injections recovered using both $\Lc$ and $\Lc^{mx}$ for various detector networks. Number of injections recovered at higher distance (i.e low SNR injection) are increasing with increase in number of detectors. Also, at higher distance,  $\Lc^{mx}$ is recovering more injections compared to $\Lc$. This is because as distance increases the SNR of injections decreases. As the threshold value $\Lc^{mx}$ statistic is less than that of $\Lc$, the improvement in injection recovery of $\Lc^{mx}$ over $\Lc$ increases with distance.

In Figure \ref{Chap4.Drate_Epsi_2} left panel,  we draw  the histograms of DNS injections exclusively recovered by $\Lc^{mx}$ (missed by $\Lc$)  with respect to  luminosity distance for  representative 2, 3, 4 and 5 detector networks LH, LHV, LHVK and LHVKI. One can see the that average luminosity distance of recovered injections increases with number of detectors in the network. The injections exclusively recovered by $\Lc^{mx}$ are the injections with SNR just above the threshold value. The network SNR of an injection increases with number of detectors in the network.  Therefore, with more detectors in the network,  the network will be able to see deep in sky.

In Figure \ref{Chap4.Drate_Epsi_2} right panel,  we draw  the histograms of DNS injections exclusively recovered by $\Lc^{mx}$ (missed by $\Lc$)  with respect to  luminosity distance  for representative 2, 3, 4 and 5 detector networks LH, LHV, LHVK, and LHVKI.
The inclination angle distributions of injections exclusively recovered by $\Lc^{mx}$ shows a trend very similar to the inclination angle of detected events proposed in \cite{Schutz:2011tw} (see Equation.28 and Figure 4). 
If all the other binary parameters are fixed, the network SNR is maximum for $\epsilon = 0, 180^\circ$ (face-on/off) and minimum for $\epsilon = 90^\circ$ (edge-on). The injections are sampled with uniform  distribution of $\cos \epsilon$, which gives less number of near face-on injections as compared to edge-on. Due to the combined effect of SNR variation and the distribution of injections,  the recovered injection follows the $\epsilon$ distribution in Figure \ref{Chap4.Drate_Epsi_2} (right panel).

\subsection{Variation in injection recovery with spin of BH component}
\begin{figure}[ht]
	\centering  	\includegraphics[width=0.8\textwidth]{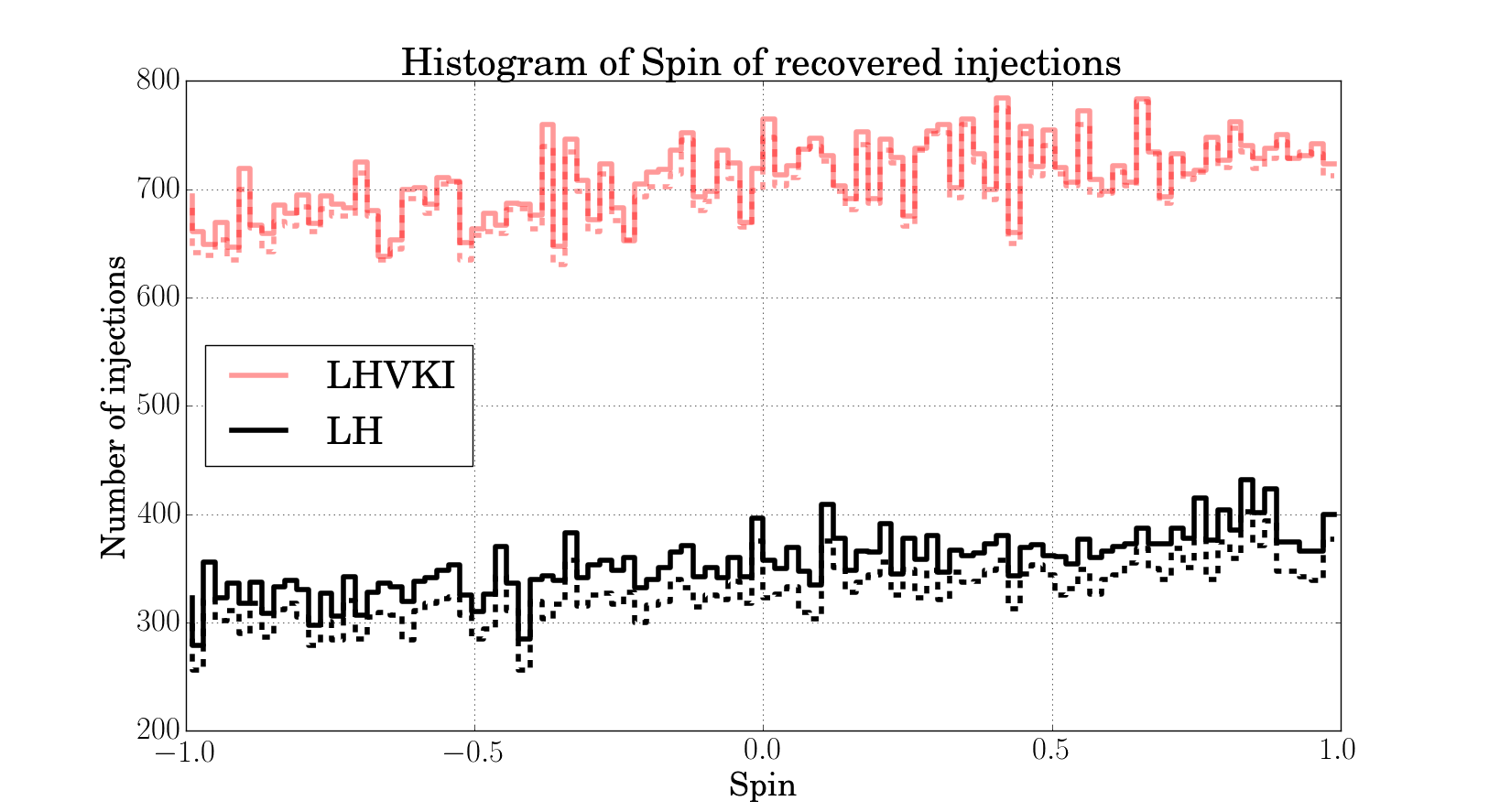}
	
	\caption{Distribution of spin of injections recovered using  statistic $\Lc$ (dashed line) and $\Lc^{mx}$ (solid line) for different detector networks. }   \label{Chap4.Rec_Spin_Hist}
\end{figure}

From Table \ref{Chap4.DetectionRate5NS_BH} it can be seen that there is an over all small increase in injection recovery for non-spinning NS-BH system compared to spinning NS-BH system for both $\Lc$ and $\Lc^{mx}$. This is because of the fact that on an average  the non-spinning injections have higher SNR compared to spinning injections.  This increase in injection recovery becomes less relevant as we move to more number of detectors. Figure \ref{Chap4.Rec_Spin_Hist} shows the histogram of spin of  injections recovered using $\Lc$ and $\Lc^{mx}$ for two different networks.

\section{Conclusion}\label{conclusions}

In this paper we perform quantitative performance comparison  between the multi-detector MLR statistic $\Lc$ and the {\it hybrid} statistic $\Lc^{mx}$ for various GW detector networks. We compute and compare the signal recovery rates of $\Lc$ and $\Lc^{mx}$ for a given fixed \FAP$=10^{-7}$.

We have demonstrated the performance by using the noise model as Gaussian with  \enquote{zero-detuning, high power} Advanced LIGO PSD \cite{aLIGOSensitivity} for a $(1.4 - 10)M_\odot$ NS-BH system with spinning/non-spinning BH and a DNS system with masses a $(1.4 - 1.4)M_\odot$. The ROC curves are used as a tool for this demonstration. The source location is sampled uniformly from the sky sphere with the distance  range $100$  \Mpc - $1$  Gpc. The inclination angle and polarization angle are sampled uniformly from the polarization sphere. The ROC curves are generated for representative 2, 3, 4 and 5 detector combinations.  On average, the generic MLR statistic recovers  $5.587\%$ ($34.45\%$),  $9.917\%$ ($49.69\%$),  $14.73\%$ ($61.25\%$) and  $19.86\%$ ($69.67\%$) of DNS (NS-BH) injections  for 2, 3, 4 and 5 detector networks respectively for a fixed \FAP of  $10^{-7}$.   The {\it hybrid} statistic shows $\sim 14.63\%$ ($7.45\%$), $\sim 12.91\%$ ($4.57\%$), $\sim 11.49\%$ ($2.56\%$) and $\sim 10.29\%$ ($1.22\%$) relative improvement in the injection recovery  rate over generic MLR statistic for 2, 3, 4 and 5 detector networks respectively for DNS (NS-BH) injections. 

The real GW detector noise is not pure Gaussian but is contaminated by  the non-Gaussian noise transients namely glitches.
Hence for real searches, one might need to add weight to $\Lc^{mx}$ using the $\chi^2$ statistic in the similar way demonstrated in \cite{Harry:2010fr} for $\Lc$. 

In this study, all the  simulations are carried out for  binary systems with fixed masses. In real search scenario, the masses are unknown and	 one needs to place templates in mass space and perform the search. A template-based search increases the false alarms. However, this applies to the search based on both {\it hybrid} statistic $\Lc^{mx}$ and the MLR statistic $\Lc$. Further owing to a single stream, we expect to get less false alarms for {\it hybrid} statistic as compared to the MLR statistic.

In  \cite{Williamson:2014wma}, authors  used a face on/off tuned MLR statistic (single stream) for the  GW follow-up search of short Gamma Ray Bursts (GRBs) of IPN triggers.  This was targeted search with templates in mass parameter space in LIGO-Virgo data. The results did show a similar improvement in the false alarm rates compared to generic MLR statistic. The short GRBs are expected to have narrow opening angle $(<30^\circ)$ and the IPN search is carried out with small inclination angles. However, in this paper we show that {\it hybrid} statistic can preform better than the generic MLR statistic for much wider region of inclination angle and hence it can be used for the  search of binary inspirals having arbitrary inclination angle without compromising too much on the network SNR for different combinations of networks.

\section{Acknowledgment}
The authors availed with the 128 cores computing facility established by the MPG-DST Max Planck Partner Group  at Indian Institute of Science Education and Research Thiruvananthapuram. Vinaya Valsan is supported by  INSPIRE program of  Department of Science and Technology, India.  The authors would like to thank Collin Capano for reviewing the draft and for useful comments. This document has been assigned LIGO laboratory document number LIGO-P1600331.

\bibliographystyle{ws-ijmpd}
\bibliography{GWref}

\end{document}